%% file: CDC2022.tex
\documentclass[letterpaper, 10 pt, conference]{ieeeconf}  

\IEEEoverridecommandlockouts                              

\usepackage{comment}

\usepackage{amsmath}
\usepackage{amsfonts}

\newtheorem{theorem}{Theorem}
\newtheorem{remark}{Remark}
\newtheorem{problem}{Problem}

\newcommand{\cfunof}[1]{\ensuremath{\left\{#1\right\}}}
\newcommand{\funof}[1]{\ensuremath{\left(#1\right)}}
\newcommand{\norm}[1]{\ensuremath{\left\Vert #1 \right\Vert}}
\newcommand{\R}{\ensuremath{\mathbb{R}}}
\newcommand{\s}{\ensuremath{\left(s\right)}}
\newcommand{\trace}{\ensuremath{\mathrm{tr}\,}}

\usepackage{graphicx}
\usepackage{subcaption}
\usepackage{pgfplots}
    \pgfplotsset{
        colormap={parula}{
            rgb255=(53,42,135)
            rgb255=(15,92,221)
            rgb255=(18,125,216)
            rgb255=(7,156,207)
            rgb255=(21,177,180)
            rgb255=(89,189,140)
            rgb255=(165,190,107)
            rgb255=(225,185,82)
            rgb255=(252,206,46)
            rgb255=(249,251,14)
        },
    }
    \pgfplotsset{compat=newest}
    \pgfplotsset{
    minor grid style={dashed,red}
    }

\usetikzlibrary{spy,shapes}

\pgfdeclarelayer{fg}    
\pgfsetlayers{main,fg}  

\usepackage[utf8]{inputenc}
\usepackage[T1]{fontenc}
\pgfplotsset{compat=1.15}

\pgfplotsset{compat=newest} 
\usepackage{tikz}
\usetikzlibrary{arrows,shapes,positioning}

\makeatletter
\let\NAT@parse\undefined
\makeatother

\usepackage{hyperref}
\usepackage[nameinlink]{cleveref}
\usepackage{autonum}

\definecolor{lightblue}{rgb}{0.54, 0.81, 0.94}
\hypersetup{
    colorlinks = true,
    linkcolor = lightblue,
    citecolor = lightblue,
    urlcolor = lightblue,
    }

\crefformat{equation}{(#2#1#3)}
\crefmultiformat{equation}{(#2#1#3)}%
{ and~(#2#1#3)}{, (#2#1#3)}{ and~(#2#1#3)}
\crefname{problem}{problem}{problems}

\tikzset{every picture/.style=thick}

\title{\LARGE \bf
Fundamental limitations on the control of lossless systems
}

\author{Johan Lindberg and Richard Pates
\thanks{The authors are with the Department of Automatic Control, Lund University, Box 118, {SE-221 00 Lund}, Sweden. They are members of the ELLIIT Strategic Research Area at Lund University. This work was supported by the ELLIIT Strategic Research Area. This project has received funding from ERC grant agreement No 834142. e-mail: johan.lindberg@control.lth.se, richard.pates@control.lth.se}
}

\begin{document}

\maketitle
\thispagestyle{empty}
\pagestyle{empty}

\begin{abstract}
In this paper we derive fundamental limitations on the levels of $H_2$ and $H_\infty{}$ performance that can be achieved when controlling lossless systems. The results are applied to the swing equation power system model, where it is shown that the fundamental limit on the $H_2$ norm scales with the inverse of the harmonic mean of the inertias in the system. This indicates that power systems may see a degradation in performance as more renewables are integrated, further motivating the need for new control solutions to aid the energy transition.
\end{abstract}

\textit{\textbf{Keywords - }}\textbf{Optimal control, Power systems}


\section{Introduction}

The lossless systems form an important class of models. They are frequently used to explain and understand phenomena arising in engineering applications. This is particularly true when describing the transportation of physical quantities, such as electrical power. This is because it is typically desirable to engineer such systems to minimise losses, making the resulting dynamical systems amenable to modelling within the lossless framework. Furthermore lossless systems enjoy rich theoretical properties. For example, factorisations involving lossless transfer functions play a crucial role in $H_\infty{}$ methods \cite{Kim97}. In addition, central control theoretic results, such as the Kalman-Yakubovich-Popov Lemma, simplify significantly in the lossless setting \cite{Wil72}, and the state-space and circuit theoretic descriptions of lossless systems have a range of appealing structural properties \cite{Hug17c,Pat22}. 

In this article, we study the following optimal control problem:

\begin{problem}\label{prob:1}
Let 
\begin{equation}\label{eq:p1}
\begin{aligned}
\dot{x}&=Ax+B\funof{u+w_u},\,x\funof{0}=0,\\
z&=\begin{bmatrix}
C&D\\0&I
\end{bmatrix}
\begin{bmatrix}
x\\u
\end{bmatrix}\\
y&=Cx+D\funof{u+w_u}+w_y,
\end{aligned}
\end{equation}
and
\begin{equation}\label{eq:p2}
\begin{aligned}
\dot{x}_{\mathrm{K}}&=A_{\mathrm{K}}x_{\mathrm{K}}+B_{\mathrm{K}}y,\,x_{\mathrm{K}}\funof{0}=0,\\
u&=C_{\mathrm{K}}x_{\mathrm{K}}+D_{\mathrm{K}}y,
\end{aligned}
\end{equation}
and denote the closed-loop transfer function from $w=\begin{bmatrix} w_u^{\mathsf{T}} & w_y^{\mathsf{T}} \end{bmatrix}^{\mathsf{T}}$ to $z$ as defined by \cref{eq:p1,eq:p2} as $T_{zw}\s$. Find
\[
\gamma^*_{\bullet}=\inf\cfunof{\gamma:\norm{T_{zw}\s}_{\bullet}<\gamma},
\]
where $\norm{\cdot}_{\bullet}$ denotes either the $H_2$ or $H_{\infty}$ norm.
\end{problem}

This is a standard setup, in which the objective is to design a dynamic feedback control law
\[
K\s=C_{\mathrm{K}}\funof{sI-A_\mathrm{K}}^{-1}B_\mathrm{K}+D_{\mathrm{K}}
\]
to minimise the effects of process disturbances and sensor noise on the output and control effort of a process with dynamics
\begin{equation}\label{eq:process}
G\s=C\funof{sI-A}^{-1}B+D.
\end{equation}
The main theoretical contribution, given as \Cref{thm:main} in \Cref{sec:2}, is to show that when the process $G\s$ is lossless, $\gamma^*_{H_2}=\sqrt{2\,\trace{\funof{CB}}}$ and if in addition $D=0$, $\gamma^*_{H_\infty{}}=\sqrt{2}$.

The utility of this result stems from the fact that it analytically characterises fundamental limitations on the control of lossless systems. For example, since electric power systems at the transmission and sub-transmission level are close to lossless, it shows that \textit{no matter how they are designed}, power system controllers in this part of the grid can never achieve better levels of $H_2$ or $H_\infty{}$ performance than $\gamma^*_{H_2}$ and $\gamma^*_{H_\infty}$. Since the derived expressions are analytical, they can be rewritten in terms of the model parameters. In \Cref{sec:3a} we use this to highlight that system inertia plays a fundamental role in the control of power systems, by showing that $\gamma^*_{H_2}$ scales with the inverse of the harmonic mean of the inertias in the system. As discussed in \Cref{sec:3b}, this suggests that the decrease in system inertia and increase in stochastic disturbances that accompanies the introduction of renewables \cite{Jor18} can significantly deteriorate performance, as quantified by the $H_2$ norm. This provides further evidence that new control approaches are required to support the energy transition, perhaps through the use of more advanced measurement tools \cite{BTN14}.

\begin{figure}
\centering
\begin{tikzpicture}[>=stealth]

  \node[rectangle, draw, minimum width=1.25cm, minimum height=1cm] (K) at
(0,0) {$K\left(s\right)$};
  \node[rectangle, draw, minimum width=1.25cm, minimum height=1cm] (G) at
(3,0) {$G\left(s\right)$};
  \node[circle, draw, inner sep=0] (plus1) at (1.5,0) {$+$};
  \node[circle, draw, inner sep=0] (plus2) at (4.5,-1.5) {$+$};
  \coordinate (end) at (5.5,0);

  \draw[->] (plus2.west) -| ([xshift=-1cm]K.west) -- node[above] {$y$} (K.west);
  \draw[->] ([yshift=-1.5cm]end) -- node[above, xshift=0.2cm] {$w_y$}
(plus2.east);
  \draw[->] (K.east) -- node[above] {$\phantom{y}u\phantom{y}$} (plus1.west);
  \draw[->] ([yshift=.75cm]plus1.north) -- node[right] {$w_u$} (plus1.north);
  \draw[->] (plus1.east) -- (G.west);
  \draw[->] (G.east) -- ([xshift=-1cm]end) -- node[above] {$y_{\mathrm{G}}$} (end);
  \draw[->] ([yshift=1.5cm]4.5,-1.5) -- (plus2.north);

\end{tikzpicture}
\caption{\label{fig:prob1}Block diagram representation of \Cref{prob:1}. The objective is to design a feedback control law to minimise the effects of process disturbances and sensor noise ($w_u$ and $w_y$) on the output and control effort ($y_{\mathrm{G}}$ and $u$) of a process.}
\end{figure}
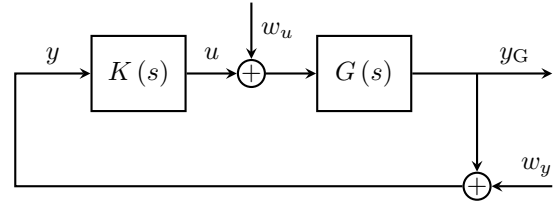

These results naturally complement existing results on fundamental limitations on the control of large-scale systems. There the focus has typically been on the performance limits imposed by restrictions on controller structure, such as locality \cite{BPD02}. However, due to the difficulty of the underlying mathematical problems \cite{Wit68,LL11}, there are few extensions of these results that cover broader classes of controller dynamics \cite{TS17}. In contrast, the limitations derived here hold for all causal controllers. Although no locality restrictions are imposed, the optimal control laws associated with \Cref{prob:1} are inherently structured. This is discussed in \Cref{sec:3c}, where it is demonstrated that while the optimal control laws cannot prevent the emergence of undesirable behaviours, they can prevent them spreading throughout the system.

\section{Fundamental limitations}\label{sec:2}

A transfer function $G\s$ in the form of \cref{eq:process} is said to be lossless if $G\s=-G\funof{-s}^{\mathsf{T}}$. Other equivalent descriptions of losslessness include the condition given as \cref{eq:lossless} below, and that the state-space model $\dot{x}=Ax+Bu$, $y=Cx+Du$ defines a (behavioral) description of the driving point behavior of an electrical network constructed with only capacitors, inductors, transformers and gyrators \cite{Hug17c}. The following theorem shows that if the process to be controlled \cref{eq:process} is lossless (and has no direct term in the $H_\infty$ case), then both $\gamma_{H_2}^*$ and $\gamma_{H_\infty{}}^*$ can be determined analytically. These expressions thus impose fundamental limits on the levels of $H_2$ and $H_\infty{}$ performance that can be achieved when controlling lossless systems. These limitations will be interpreted in the context of simple electric power system models in the next section.

\begin{theorem}\label{thm:main}
Assume that $A$, $B$, $C$, and $D$ are as in \Cref{prob:1}, and that the pair $\funof{A,B}$ is controllable. If there exists a positive definite $P$ such that
\begin{equation}\label{eq:lossless}
PA+A^{\mathsf{T}}P=0,\;PB=C^{\mathsf{T}},\;\text{and}\;D+D^{\mathsf{T}}=0,
\end{equation}
then in the $H_2$ case of \Cref{prob:1}, 
\[
\gamma^*_{H_2}=\sqrt{2\,\trace{\funof{CB}}}.
\]
If in addition $D=0$, then $\gamma^*_{H_\infty{}}=\sqrt{2}$.
\end{theorem}

\begin{remark}
It is shown in \cite{Hug17c} that the dynamics of lossless systems are always behaviorally controllable, meaning that the controllability assumption in \Cref{thm:main} is essentially without loss of generality.
\end{remark}

\begin{proof}
The dynamics in \cref{eq:p1} are a special case of the generalised plant dynamics
\begin{equation}\label{eq:genpl}
\begin{aligned}
\begin{bmatrix}
\dot{x}\\z\\y
\end{bmatrix}
=
\begin{bmatrix}
A&B_w&B\\C_z&0&D_{zu}\\C&D_{yw}&D_{yu}
\end{bmatrix}
\begin{bmatrix}
x\\w\\u
\end{bmatrix},
\end{aligned}
\end{equation}
where
\[
B_w=\begin{bmatrix}B&0\end{bmatrix},\,
C_z=\begin{bmatrix}C\\0\end{bmatrix},
\]
and
\[
D_{zu}=\begin{bmatrix}D\\I\end{bmatrix},\,
D_{yw}=\begin{bmatrix}D&I\end{bmatrix},\,D_{yu}=D.
\]
Observe that under the conditions of the theorem statement, the pairs $\funof{A,B_w}$ and $\funof{A,B}$ are controllable. To see this, first note that controllability of $\funof{A,B}$ implies controllability of $\funof{A,B_w}$. Since $P$ is invertible, controllability of \funof{A,B} implies that $\funof{-PAP^{-1},PB}$ is also controllable. Therefore $\funof{B^{\mathsf{T}}P,-P^{-1}A^{\mathsf{T}}P}$ is observable, which implies that $\funof{C,A}$ is observable by \cref{eq:lossless}. Observability of $\funof{C,A}$ then implies observability of $\funof{C_z,A}$. Furthermore the matrices $D_{zu}$ and $D_{yw}$ are full rank. Therefore the $H_2$ and $H_{\infty}$ solutions to \Cref{prob:1} can be tackled within the Riccati equation framework (see for example the requirements on p.383--384 of \cite{ZDG96}). We will first prove the result under the simplifying assumption that $D=0$. In this case the generalised plant in \cref{eq:genpl} meets the conditions of \cite[\S{III.A}]{DKGF89}. This simplifies the Riccati equations that must be solved significantly, and illustrates the key steps of the proof. We will then show how to extend the approach for $D\neq{}0$ in the $H_2$ case.

\textit{The $H_2$ case:} Let $X$ denote the unique stabilising solution of
\begin{equation}\label{eq:h21}
XA+A^{\mathsf{T}}X-XBB^{\mathsf{T}}X+C_z^{\mathsf{T}}C_z=0,
\end{equation}
and $Y$ denote the unique stabilising solution of
\begin{equation}\label{eq:h22}
YA^{\mathsf{T}}+AY-YC^{\mathsf{T}}CY+B_wB_w^{\mathsf{T}}=0.
\end{equation}
By \cite[Theorem 1]{DKGF89}, $
\gamma^*_{H_2}=\sqrt{\norm{G_\mathrm{ctl}\s}_{H_2}^2+\norm{G_\mathrm{obs}\s}_{H_2}^2}
$, where
\[
\begin{aligned}
G_\mathrm{ctl}\s&=\funof{C_z-D_{zu}B^{\mathsf{T}}X}\funof{sI-A+BB^{\mathsf{T}}X}^{-1}B_w,\\
G_\mathrm{obs}\s&=B^{\mathsf{T}}X\funof{sI-A+YC^{\mathsf{T}}C}^{-1}\funof{B_w-YC^{\mathsf{T}}D_{yw}}.
\end{aligned}
\]
By \cite[Corollary 13.8]{ZDG96} $X$ and $Y$ correspond to the unique positive-semidefinite solutions to the given Riccati equations. Since
$B_wB_w^{\mathsf{T}}=BB^{\mathsf{T}}$ and $C_z^{\mathsf{T}}C_z=C^{\mathsf{T}}C$, we then see by comparison with \cref{eq:lossless} that
\begin{equation}\label{eq:solh2}
X=P\;\;\text{and}\;\;Y=P^{-1}.
\end{equation}
Substituting the above into the expressions for $G_\mathrm{ctl}\s$ and $G_\mathrm{obs}\s$ shows that
\[
\begin{aligned}
G_\mathrm{ctl}\s&=
\begin{bmatrix}C\\-C\end{bmatrix}
\funof{sI-A+BC}^{-1}\begin{bmatrix}B&0\end{bmatrix},\\
G_\mathrm{obs}\s&=C\funof{sI-A+BC}^{-1}
\begin{bmatrix}B&-B\end{bmatrix}.
\end{aligned}
\]
Therefore
\[
\gamma^*_{H_2}=\sqrt{2\norm{G_\mathrm{obs}\s}_{H_2}^2}=\sqrt{2\,\trace{\funof{CZC^{\mathsf{T}}}}},
\]
where $Z$ is the solution to the Lyapunov equation
\[
\funof{A-BC}Z +Z \funof{A^{\mathsf{T}}-C^{\mathsf{T}}B^{\mathsf{T}}} +2B B^{\mathsf{T}}=0.
\]
By comparison with \cref{eq:lossless} we see that $Z=P^{-1}$, which implies that $\gamma^*_{H_2}=\sqrt{2\,\trace{\funof{C P^{-1} C^{\mathsf{T}}}}}=\sqrt{2\,\trace{\funof{CB}}}$ as required.

\textit{The $H_{\infty}$ case:} Whenever such solutions exist, let $X_\gamma$ denote the unique stabilising solution of
\[
X_\gamma{}A+A^{\mathsf{T}}X_\gamma{}-X_\gamma{}\funof{BB^{\mathsf{T}}-\gamma^{-2}B_wB_w^\mathsf{T}}X_\gamma{}+C_z^{\mathsf{T}}C_z=0,
\]
and $Y_\gamma$ denote the unique stabilising solution of
\[
Y_\gamma{}A^{\mathsf{T}}+AY_\gamma{}-Y_\gamma{}\funof{C^{\mathsf{T}}C-\gamma^{-2}C_z^\mathsf{T}C_z}Y_\gamma{}+B_wB_w^{\mathsf{T}}=0.
\]
By \cite[Theorem 3]{DKGF89}, $\norm{T_{zw}}_{H_\infty}<\gamma$ if and only if $X_\gamma$ and $Y_\gamma$ exist, and the magnitude of the largest eigenvalue of $X_\gamma{}Y_\gamma$ is less than $\gamma^2$. By \cite[Corollary 13.8]{ZDG96}, if $\gamma>1$ then $X_\gamma{}$ and $Y_\gamma{}$ exist, and correspond to the unique positive-semidefinite solutions to the given Riccati equations. Comparison with \cref{eq:lossless} shows that
\[
X_{\gamma}=\tfrac{\gamma}{\sqrt{\gamma^2-1}}P,\;\;\text{and}\;\;Y_{\gamma}=\tfrac{\gamma}{\sqrt{\gamma^2-1}}P^{-1}.
\]
Therefore the magnitude of the largest eigenvalue of $X_\gamma{}Y_\gamma$ equals $\gamma^2/\funof{\gamma^2-1}$, which implies that $\gamma^*_{H_\infty{}}=\sqrt{2}$.

\textit{The case $D\neq0$:} Loop shifting can be used to extend the above arguments to allow for nonzero $D$. As explained in \cite[p.453--454]{ZDG96}, the $H_2$ optimal control problem with generalised plant as in \cref{eq:genpl} is equivalent to the $H_2$ optimal control problem with generalised plant
\[
\begin{bmatrix}
\dot{x}\\\tilde{z}\\\tilde{y}
\end{bmatrix}
=
\begin{bmatrix}
A&B_w&BS\\C_z&0&\begin{bmatrix}D\\I\end{bmatrix}S^{-1}\\RC&R^{-1}\begin{bmatrix}
D&I
\end{bmatrix}&0
\end{bmatrix}
\begin{bmatrix}
x\\\tilde{w}\\\tilde{u}
\end{bmatrix},
\]
where $R=\funof{I+DD^\mathsf{T}}^{\frac{1}{2}}$ and $S=\funof{I+D^\mathsf{T}D}^{\frac{1}{2}}$. This generalised plant meets the conditions on \cite[p.384]{ZDG96}. Therefore $\gamma_{H_2}^*$ can be determined using arguments based on Riccati equations as described in the $D=0$ case, but with the Riccati equations in \cref{eq:h21,eq:h22} replaced with their `generalised' counterparts, as specified in \cite[Theorem 14.7]{ZDG96}. These also admit the solutions in \cref{eq:solh2}, resulting in the same optimal value for $\gamma_{H_2}^*$. 
\end{proof}

\section{Implications for power system control}\label{sec:3}

In this section we present and discuss the application of \Cref{thm:main} to simple models of electric power systems at the transmission and sub-transmission level. These results, although preliminary since they lack the support of a detailed numerical study from a realistic power system model, clearly point to the fundamental role played by inertia when considering the control of power systems.

\subsection{Applying \Cref{thm:main} to swing equation models}
\label{sec:3a}

In the absence of damping, after linearisation, the swing equation power system model is described by the equations
\begin{equation}\label{eq:swing}
\begin{aligned}
M_k\ddot{\theta}_k&=p_{\mathrm{N},k}+u_k+w_{u,k},\,\begin{bmatrix}
\theta_k\funof{0}\\\dot{\theta}_k\funof{0}
\end{bmatrix}=0,\,k\in\cfunof{1,\ldots{},n},\\
\begin{bmatrix}
p_{\mathrm{N}}\\0
\end{bmatrix}
&=-
\begin{bmatrix}
K_{\mathrm{a}}&K_{\mathrm{b}}\\
K_{\mathrm{b}}^{\mathsf{T}}&K_{\mathrm{c}}
\end{bmatrix}\begin{bmatrix}
\theta\\\theta_{\mathrm{int}}
\end{bmatrix}.
\end{aligned} 
\end{equation}
In the above, the first equation describes the dynamics at a set of $n$ generator buses. $M_k>0$ denotes the inertia parameter of the \emph{k}th generator bus, $p_{\mathrm{N},k}$ the power injection from the transmission network, $u_k$ an adjustable power injection, and $w_{u,k}$ a power disturbance. The variable $\theta$ gives the vector of electrical angles at the generator buses, and $\theta_{\mathrm{int}}$ the angles at the remaining buses. The second equation describes how these relate to the power injections $p_{\mathrm{N}}$ as defined by the dynamics of the transmission network. In particular the block matrix with entries $K_{\mathrm{a}}\in\R^{n\times{}n}$, $K_{\mathrm{b}}\in\R^{n\times{}m}$ and $K_{\mathrm{c}}\in\R^{m\times{}m}$, is a weighted Laplacian matrix, with edge weights determined by the line susceptances and load angles across the lines at equilibrium. 

We consider the problem of how to select the adjustable power injections $u$, when a set of noisy angular frequency measurements
\begin{equation}\label{eq:thetm}
y_k=\dot{\theta}_k+w_{y,k},\;k\in\cfunof{1,\ldots{},n},
\end{equation}
where $w_{y,k}$ denotes the measurement noise, are available. In the remainder of this subsection, we will show that under a set of mild assumptions this problem fits naturally into the framework of \Cref{prob:1}. Furthermore, for an appropriate choice of the system state, all the conditions of \Cref{thm:main} are satisfied, and the expression for $\gamma_{H_2}^*$ simplifies further.

The required assumptions are:
\begin{enumerate}
    \item[A1] The transmission network is connected.
    \item[A2] The load angle across every transmission line has magnitude less than $90^\circ$.
\end{enumerate}
A1 is essentially without loss of generality, since if the network is not connected, every connected component can be analysed separately. A2 is required to ensure the weighted edges in the Laplacian are positive. A2 will likely be satisfied in practice, since operational limits on transmission lines typically preclude load angles greater than $45^\circ$ \cite{Kun94}[p.230].

Under A1--A2, the matrix $K_{\mathrm{red}}=K_{\mathrm{a}}-K_{\mathrm{b}}K_{\mathrm{c}}^{-1}K_{\mathrm{b}}^{\mathsf{T}}$ can be factored as $K_{\mathrm{red}}=LL^\mathsf{T}$, where $L\in\R^{n\times{}n-1}$ has full (column) rank. Letting $M\in\R^{n\times{}n}$ denote the diagonal matrix with \emph{k}th diagonal entry equal to $M_k$ and $x=\begin{bmatrix}\dot{\theta}^\mathsf{T}&\theta{}^{\mathsf{T}}L\end{bmatrix}^{\mathsf{T}}$, it then follows that \cref{eq:swing,eq:thetm} can be written as
\begin{equation}\label{eq:ssswing}
\begin{aligned}
\dot{x}&=\begin{bmatrix}
0&-M^{-1}L\\L^{\mathsf{T}}&0
\end{bmatrix}x+
\begin{bmatrix}
M^{-1}\\0
\end{bmatrix}\funof{u+w_u},\,x\funof{0}=0,\\
y&=\begin{bmatrix}I&0
\end{bmatrix}x+w_y.
\end{aligned}
\end{equation}
These equations take the form of the first and third equations in \cref{eq:p1}. Therefore in the context of the swing equation power system model, \Cref{prob:1} corresponds to searching for a control law to minimise the effects of power disturbances and measurement noise on the deviations in electrical frequency and control effort. 

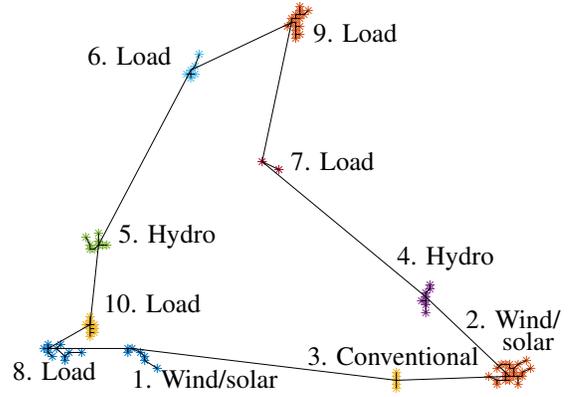
\begin{figure}
    \centering
    \input{Tikz/Network}
    
    \vspace{-1cm}
    
    \caption{Example of a network with 100 buses.}
    \label{fig:network}
\end{figure}

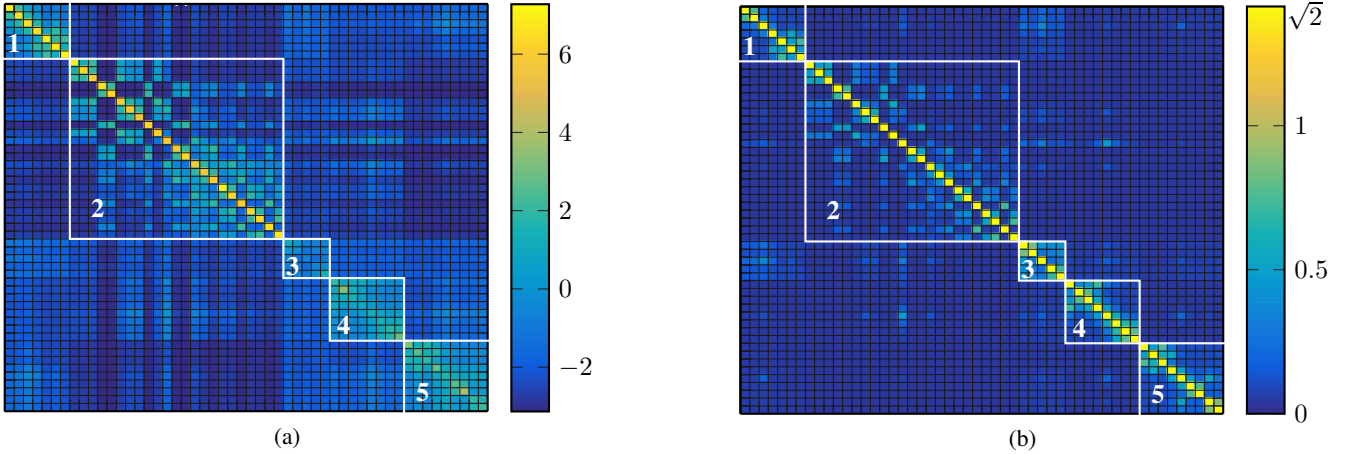
\begin{figure*}
\vspace{.2cm}
    \centering
    \begin{subfigure}{0.45\textwidth}
        \centering
        \input{Tikz/H2}
        
        \vspace{-.7cm}
        
        \caption{}
        \label{fig:H_2}
    \end{subfigure}%
    \hfill{}
    \begin{subfigure}{0.45\textwidth}
        \centering
        \input{Tikz/H_inf}
        
        \vspace{-.7cm}
        
        \caption{}
        \label{fig:H_inf}
    \end{subfigure}%
\caption{(a)--(b) $\ln{(\gamma_{H_2,ik})}$ and $\gamma_{H_{\infty},ik}$, respectively, for the power systems model in \Cref{fig:network}.}
\end{figure*}

Setting 
\[
P=\begin{bmatrix}
M&0\\0&I
\end{bmatrix}
\]
shows that the conditions of \Cref{thm:main} are satisfied. 
The controllability assumption is satisfied since the controllability matrix equals
\[
\begin{bmatrix}
M^{-1}&0&\cdots{}\\
0&L^{\mathsf{T}}M^{-1}&\cdots{}
\end{bmatrix}\in\R^{\funof{2n-1}\times{}n\funof{2n-1}},
\]
which has rank equal to $2n-1$ (both $M$ and $L$ have full rank). Furthermore, the expression for the $H_2$ norm in \Cref{thm:main} simplifies to
\begin{equation}\label{eq:massform}
\gamma^*_{H_2}=\sqrt{2\funof{\frac{1}{M_1}+\cdots{}+\frac{1}{M_n}}}.
\end{equation}
This implies that $\gamma^{*2}_{H_2}\,/\,n=2\,/\,\mathrm{HM}\funof{M_1,\ldots{},M_n}$, where $\mathrm{HM}$ denotes the harmonic mean. This indicates that the fundamental limit on $H_2$ performance scales with the inverse of the harmonic mean of the inertias in the system.

\begin{remark}
The lossless assumption is reasonably well justified in the power system context when considering generation and consumption at the transmission and sub-transmission level (at least from the perspective of control system design, where qualitative, simple modelling is often more appropriate). Observe also that damping effects that arise from control actions are captured by \cref{eq:swing}, since this equation includes adjustable power injections (this implies for example that droop-control cannot achieve better performance that the levels specified by \Cref{thm:main}, since a droop controller is a special case of \cref{eq:p2}).

Nevertheless, the modelling setup considered here is highly simplified. so the given expressions should be used to provide insights, and supplement other analysis approaches. In particular neglecting voltage dynamics (as is implicit in any analysis based on the swing equations) is questionable. That said, high fidelity power system models seem to be close to lossless. For example the resistances in the equivalent circuit descriptions of synchronous machines and transmission lines in \cite[Chapters 5--6]{Kun94} are very small (for reasons of space this line won't be pursued further here, it is interesting to think how \cref{eq:massform} would generalise if more sophisticated models were used). It should also be stressed that the model considered is not suitable for distribution networks where losses can be significant. 
\end{remark}

\subsection{The impact of increasing renewable generation}
\label{sec:3b}

In the power system context, the increased use of renewables is typically associated with a reduction in system inertia, an increase in stochastic disturbances, and a larger number of different components. In this subsection we interpret the effects of these trends within the context of \cref{eq:massform}. These results apply under the tacit assumption that frequency measurements are being used for control (viz. \cref{eq:thetm}). This is the status quo in practice. However it is being increasingly recognised that different approaches are required to handle the renewable transition. One possible interpretation of the results in this section is to provide further support for this, by revealing the presence of fundamental performance limitations that scale poorly with the reduction and increased heterogeneity of inertia throughout a system.

\textit{Reduced inertia and increased stochastics:} When conventional power generation is replaced by renewable sources such as wind, the total inertia in a power system is reduced. This is because the synchronous machines used in conventional generation have considerable mass. In contrast, the inertia in a wind turbine is typically electrically decoupled from the grid, so contributes relatively little inertia by comparison, and even if the power electronics in a wind turbine are used to emulate the dynamics of conventional generator (by operating the turbine as a virtual synchronous machine), the level of synthetic inertia that can be realised is typically far smaller.

The effect of reducing the inertia of the generation sources can be captured by reducing the sizes of the constants $M_k$. Interestingly this does not affect $\gamma_{H_\infty}^*$, which suggests that performance with respect to worst case disturbances may not degrade. However it is easily seen from \cref{eq:massform} that reducing the value of any $M_k$ will increase $\gamma_{H_2}^*$. Since the size of the $H_2$ norm captures how stochastic disturbances are amplified (which will become more prevalent with an increased use of renewables), this suggests that power system performance may be adversely affected by the increased use of renewables. Furthermore these performance limitations depend in a fundamental way on, for example, the sizes of the virtual inertia constants that can be synthesised, emphasising the importance of larger inertia constants in attenuating stochastic disturbances.

\textit{Increased system heterogeneity:} A secondary effect of increasing the use of renewables is that the inertia parameters in \cref{eq:swing} will cover a wider range of values. Jensen's inequality implies that
\begin{equation}\label{eq:jensen}
\sum_{k=1}^n\frac{1}{M_k}\geq{}\sum_{k=1}^n\frac{1}{M_{\mathrm{tot}}/n},\;\text{where}\;M_{\mathrm{tot}}=\sum_{k=1}^nM_k,
\end{equation}
and equality is achieved only if $M_1=M_2=\cdots{}=M_n$. Therefore given a fixed total amount of inertia in the system (constant $M_{\mathrm{tot}}$), the more similar the individual inertia parameters are (the closer the $M_k$'s are to $M_{\mathrm{tot}}/n$), the smaller $\gamma_{H_2}^*$ is. Conversely, the more heterogeneous the set of masses, the larger $\gamma_{H_2}^*$ becomes. This indicates that the sensitivity of the system to stochasticity may be further exacerbated as renewable sources are added purely as a result of having components with a wider range of inertia parameters present in the system.


\subsection{Optimal control structures} \label{sec:3c}

Condensing notions of system performance down to numbers such as $\gamma_{H_2}^*$ and $\gamma_{H_\infty}^*$, especially when the system in question is extremely large, rarely tells the full story. In this subsection we investigate performance of the swing equation power system model \cref{eq:ssswing} further by studying the $H_2$ and $H_\infty$ norms of the sub-matrices of $T_{zw}\s$ associated with individual disturbances and output signals. In particular, denoting
\[
\tilde{w}_k=\begin{bmatrix}w_{u,k}\\w_{y,k}\end{bmatrix}\;\;\text{and}\;\;\tilde{z}_k=\begin{bmatrix}\funof{Cx+Du}_k\\u_k\end{bmatrix},
\]
we study
\[
\gamma_{H_2,ik}=\norm{T_{\tilde{z}_i\tilde{w}_k}\s}_{H_2}\;\;\text{and}\;\;\gamma_{H_\infty{},ik}=\norm{T_{\tilde{z}_i\tilde{w}_k}\s}_{H_\infty{}},
\]
where $T_{\tilde{z}_i\tilde{w}_k}\s$ is the closed loop transfer function from $\tilde{w}_k$ to $\tilde{z}_i$. To do so of course requires a choice of control law. We apply the optimal control laws for \Cref{prob:1}. It turns out that in both the $H_2$ and $H_\infty{}$ case these are highly structured. More specifically (when $D=0$, as is the case in \cref{eq:ssswing}), in the $H_2$ case
\[
K\s=-C\funof{sI-A+2BC}^{-1}B
\]
is optimal for \Cref{prob:1}, and in the $H_\infty{}$ case $K\s=-\sqrt{2}I$ is optimal for \Cref{prob:1} (for a discussion on how to synthesise these types control laws in a structure exploiting manner, see \cite{Pat22}).

\textit{Model description:} In order to investigate performance, swing equation power system models were randomly generated (the code used to do this can be found at \href{https://github.com/Johan-Lindb/L-CSS22}{https://github.com/Johan-Lindb/L-CSS22}). Note that the purpose here is to investigate trends, and the numerical values obtained from the model should not be compared with results from the power system literature. Network topologies were generated by first randomly specifying the locations of 10 clusters, with a random number of generators or loads, on a map. The loads represents a collection of several loads in a distribution system and are considered to be constant. The total number of buses was selected to be 100. Within each cluster, the buses were connected through a minimum spanning tree representing a sub-transmission network. Using eigenvector centrality, the most central bus in each cluster was selected and then connected to the central buses in the other clusters. This represents the transmission network of a power system and was created to give $n-1$ contingency, which ensures that if one of the lines or clusters were taken out, the other clusters would still be connected. An example of a system that was generated is given in \Cref{fig:network}.

\begin{figure}
\vspace{.2cm}

    \centering
    \input{Tikz/H2_avr}
    \caption{The natural log of the average of $\gamma_{H_2,ik}$ for 100 different power system models with the same number of buses in each cluster.}
    \label{fig:H_2_Average}
\end{figure}
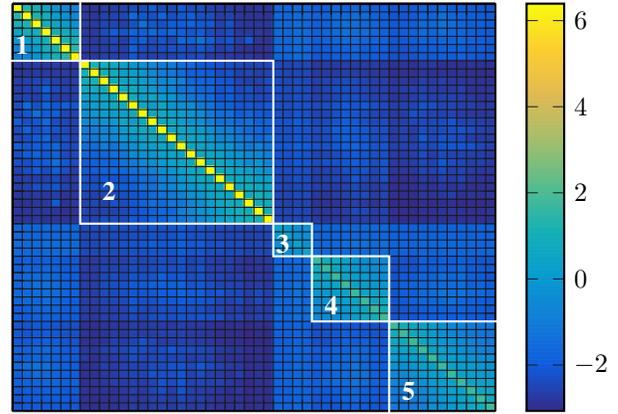

The power consumption/production of each bus was selected at random, with probability of higher magnitude in the centre of the clusters. The power producing clusters were assigned to be either conventional, hydro or wind/solar. The solar/wind generation areas are large farms with power outputs in comparison with hydro power generators or convectional generators, like large off-shore wind farms. The inertia parameter $M_k$ at each generator bus was set to be proportional to the product of the rated power of the generator, and a constant related to the type of generator. For conventional and hydro buses, these constants were chosen to be 6 and 3 respectively \cite{OKL18}. Renewables such as solar and wind typically have much lower inertias (resulting either from shunt capacitances, or designing power electronics to synthesise inertia). To reflect this, the constant for wind/solar buses was chosen to be one one-thousandth of the conventional generator constant. Other inertias were also tested. For inertias a factor of ten smaller than that of conventional and hydro power, the conclusions below still hold.

The entries of the weighted Laplacian were determined based on the line parameters of the networks. The line parameters of the lines in the sub-transmission and the transmission networks were assigned to handle all power flows, meaning that under normal operation the phase angle would be far below the 45$^\circ$ limit \cite{Kun94}, and for the transmission network also under $n-1$ contingency.  Since the reduced Laplacian $K_{\mathrm{red}}$ affects neither $\gamma^*_{H_2}$ nor $\gamma^*_{H_\infty}$, highly simplified network modelling seems sufficient.

\textit{Performance of the optimal control laws:} Consider now the network in \Cref{fig:network}, with power generation and inertia assigned to the buses as described above. \Cref{fig:H_2} shows $\ln{(\gamma_{H_2,ik})}$. Here the wind/solar generators are in the first two clusters. Observe that there are orders of magnitude differences between different values of $\gamma_{H_2,ik}$. Almost all sensitivity to the disturbances is isolated in the buses where the solar/wind generators are located. There is some disturbance sensitivity between different buses within a cluster, but between different clusters there is almost no disturbance sensitivity. This means that while a lossless power system model with renewable power generation is locally sensitive to stochastic disturbances at the buses with very little inertia, the optimal control law keeps the effect of the disturbances local. \Cref{fig:H_inf} shows $\gamma_{H_{\infty} ,ik}$. As expected from \Cref{thm:main} $\gamma_{H_{\infty,kk}}=\gamma^*_{H_{\infty}}$. Just as in the $H_2$ case, the $H_{\infty}$ optimal control law, despite being completely decentralised, keeps the effects of the disturbances mainly local.


The system in \Cref{fig:network} is just one example of a power system model. To investigate if the same conclusions would hold for other networks, 100 different networks were generated. To make the results comparable, clusters 1--2 always contained wind/solar generation, 3 conventional power generation, 4--5 hydro power generation, and 6--10 constant power loads. The sizes of the clusters were also fixed. In \Cref{fig:H_2_Average} the natural logarithm of the average of $\gamma_{H_2,ik}$ of the 100 networks is shown. The conclusions from before still hold. 
This was also true for the $H_\infty$ case, but the plot is omitted since it is very similar to \Cref{fig:H_inf}. 

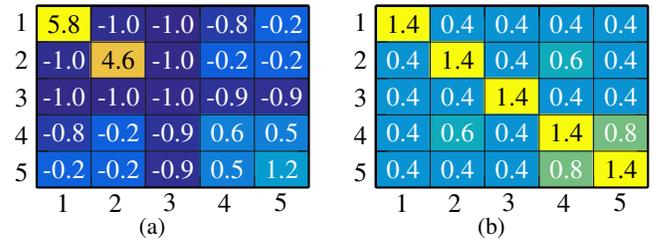
\begin{figure}
\vspace{.2cm}
    \centering
    \begin{subfigure}{0.45\columnwidth}
        \centering
        \input{Tikz/H2_red}
        
        \vspace{-.7cm}
        
        \caption{}
        \label{fig:H_2_red}
    \end{subfigure}%
    \hspace{0.5cm}
    \begin{subfigure}{0.45\columnwidth}
        \centering
        \input{Tikz/H_inf_red}
        
        \vspace{-.7cm}
        
        \caption{}
        \label{fig:H_inf_red}
    \end{subfigure}
    \caption{(a) $\ln{(\gamma_{H_2,ik})}$ for the lumped model of the power system in \Cref{fig:network}. (b) $\gamma_{H_{\infty},ik}$ for the lumped model of the power system in \Cref{fig:network}. In both subfigures, the indexing corresponds to the numbering of the clusters in \Cref{fig:network}.}
    \label{fig:H_red}
\end{figure}

\textit{Lumped models:} In power systems engineering it is very common to work with aggregated models, where several buses are lumped together. We now investigate a version of the model from the previous subsection, in which the buses within each cluster are lumped together into a single bus, with inertia and power production/consumption equal to the sum in the given cluster. The weighed Lapalacian was determined using only the lines in the transmission network. The resulting gains are shown in \Cref{fig:H_2_red,fig:H_inf_red}.

In \Cref{fig:H_2_red} it can be seen that a disturbance in a cluster affects that cluster the most. However, the $H_2$ gain is dramatically decreased from the corresponding levels in \Cref{fig:H_2}. This is a result of the heterogeneity within the cluster, as explained by \cref{eq:jensen}. The conclusion is that lumping models can give good insight into some aspects of the dynamics, but that the disturbance sensitivity of individual buses can be severely underestimated. \Cref{fig:H_inf_red} shows that lumping does not affect the performance metric when looking at the $H_{\infty}$ norm. This suggests that when looking at worst case disturbances, lumping can still give good insights. 

\section{Conclusions}

Analytical solutions to two optimal control problems involving lossless systems have been given, and interpreted in the context of electric power systems. Simplified models were used to demonstrate the presence of a fundamental performance limit that scales poorly with the model parameter changes associated with the increased use of renewables. This provides control theoretic evidence for the need for more sophisticated control methods and techniques to support the renewable energy transition.


\bibliographystyle{IEEEtran}
\bibliography{CDC2022.bib}

\end{document}

%% file: Tikz/Network.tex
  \begin{tikzpicture}
  \node[inner sep=0pt] (map) at (0,0)
    {\includegraphics[width=\linewidth]{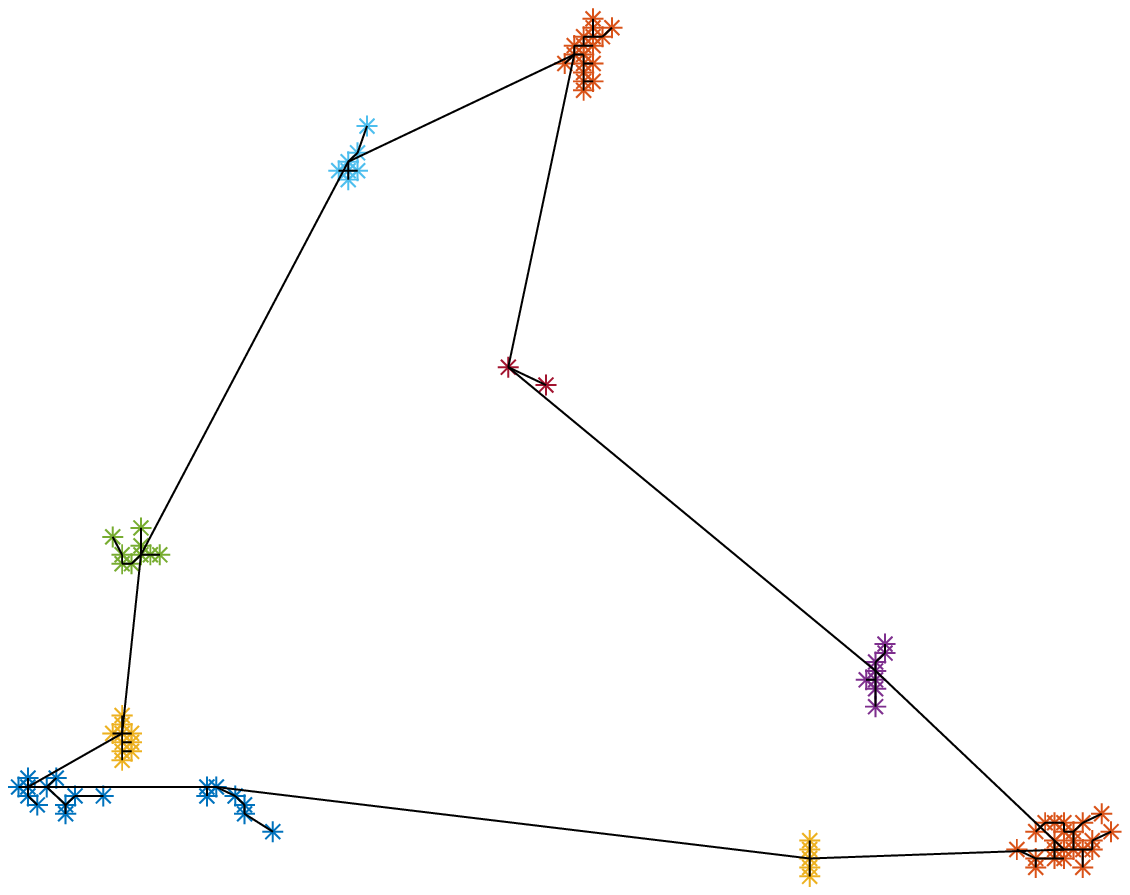}};
    \node at (-1,-2.3)[text=black] {1. Wind/solar};
    \node at (3.1, -1.5)[text=black] {2. Wind/};
    \node at (3.3,-1.8)[text=black] {solar};
    \node at (1.5,-2)[text=black] {3. Conventional};
    \node at (2.2,-0.7)[text=black] {4. Hydro};
    \node at (-1.5,-0.4)[text=black] {5. Hydro};
    \node at (-2,2)[text=black] {6. Load};
    \node at (0.7,0.6)[text=black] {7. Load};
    \node at (-3,-2.2)[text=black] {8. Load};
    \node at (1,2.3)[text=black] {9. Load};
    \node at (-1.7,-1.3)[text=black] {10. Load};
  \end{tikzpicture}

%% file: Tikz/H2.tex
\begin{tikzpicture}
    \begin{axis}[view={0}{90},
    grid=both,
    grid style={black!90},
    tick style={black!90},
    width=8cm,
    height=7cm,
        colorbar,
        y dir=reverse,
        ytick={1,2,...,52},
xtick={1,2,...,52},
yticklabels={,,},
xticklabels={,,}]
      \addplot3[surf, shader=flat corner] table {Tikz/H2.txt};
      \begin{pgfonlayer}{fg}
      \draw [draw=white, thick] (0,0) rectangle (8,8);
      \draw [draw=white, thick] (8,8) rectangle (31,31);
      \draw [draw=white, thick] (31,31) rectangle (36,36);
      \draw [draw=white, thick] (36,36) rectangle (44,44);
      \draw [draw=white, thick] (44,44) rectangle (54,54);
      
      \node at (2,6)[text=white, font=\bfseries] {1};   
      \node at (11,27)[text=white, font=\bfseries] {2};
      \node at (32,34.5)[text=white, font=\bfseries] {3};
      \node at (37.5,42)[text=white, font=\bfseries] {4};
      \node at (46,50.5)[text=white, font=\bfseries] {5};
      \end{pgfonlayer}
    \end{axis}
  \end{tikzpicture}

%% file: Tikz/H_inf.tex
\begin{tikzpicture}
    \begin{axis}[view={0}{90},
    grid=both,
    grid style={black!90},
    tick style={black!90},
    width=8cm,
    height=7cm,
        colorbar,
        y dir=reverse,
        ytick={1,2,...,52},
xtick={1,2,...,52},
yticklabels={,,},
xticklabels={,,}]
      \addplot3[surf, shader=flat corner] table {Tikz/H_inf.txt};
      \begin{pgfonlayer}{fg}
      \draw [draw=white, thick] (0,0) rectangle (8,8);
      \draw [draw=white, thick] (8,8) rectangle (31,31);
      \draw [draw=white, thick] (31,31) rectangle (36,36);
      \draw [draw=white, thick] (36,36) rectangle (44,44);
      \draw [draw=white, thick] (44,44) rectangle (54,54);
      
      \node at (2,6)[text=white, font=\bfseries] {1};   
      \node at (11,27)[text=white, font=\bfseries] {2};
      \node at (32,34.5)[text=white, font=\bfseries] {3};
      \node at (37.5,42)[text=white, font=\bfseries] {4};
      \node at (46,50.5)[text=white, font=\bfseries] {5};
      \end{pgfonlayer}
    \end{axis}
    \node at (7.5,5.3)[text=black] {$\sqrt{2}$};
  \end{tikzpicture}

%% file: Tikz/H2_avr.tex
\begin{tikzpicture}
    \begin{axis}[view={0}{90},
    grid=both,
    grid style={black!90},
    tick style={black!90},
    width=8cm,
    height=7cm,
        colorbar,
        y dir=reverse,
        ytick={1,2,...,52},
xtick={1,2,...,52},
yticklabels={,,},
xticklabels={,,}]
      \addplot3[surf, shader=flat corner] table {Tikz/H2_avr.txt};
      \begin{pgfonlayer}{fg}
      \draw [draw=white, thick] (0,0) rectangle (8,8);
      \draw [draw=white, thick] (8,8) rectangle (28,28);
      \draw [draw=white, thick] (28,28) rectangle (32,32);
      \draw [draw=white, thick] (32,32) rectangle (40,40);
      \draw [draw=white, thick] (40,40) rectangle (52,52);
      
      \node at (2,6)[text=white, font=\bfseries] {1};   
      \node at (11,24)[text=white, font=\bfseries] {2};
      \node at (29,30.5)[text=white, font=\bfseries] {3};
      \node at (34,38)[text=white, font=\bfseries] {4};
      \node at (42,48.5)[text=white, font=\bfseries] {5};
      \end{pgfonlayer}
    \end{axis}
  \end{tikzpicture}

%% file: Tikz/H2_red.tex
 \begin{tikzpicture}
    \begin{axis}[view={0}{90},
    grid=both,
    grid style={black!90},
    tick style={black!90},
    width=5.2cm,
    height=4cm,
        y dir=reverse,
        ytick={1,2,...,5},
xtick={1,2,...,5},
yticklabels={,,},
xticklabels={,,}]
      \addplot3[surf, shader=flat corner, scale=0.5] table {Tikz/H2_red.txt};
      \node at (1.5,1.5)[text=black] {5.8};
      \node at (1.5,2.5)[text=white] {-1.0};
      \node at (1.5,3.5)[text=white] {-1.0};
      \node at (1.5,4.5)[text=white] {-0.8};
      \node at (1.5,5.5)[text=white] {-0.2};
      
      \node at (2.5,1.5)[text=white] {-1.0};
      \node at (2.5,2.5)[text=black] {4.6};
      \node at (2.5,3.5)[text=white] {-1.0};
      \node at (2.5,4.5)[text=white] {-0.2};
      \node at (2.5,5.5)[text=white] {-0.2};
      
      \node at (3.5,1.5)[text=white] {-1.0};
      \node at (3.5,2.5)[text=white] {-1.0};
      \node at (3.5,3.5)[text=white] {-1.0};
      \node at (3.5,4.5)[text=white] {-0.9};
      \node at (3.5,5.5)[text=white] {-0.9};
      
      \node at (4.5,1.5)[text=white] {-0.8};
      \node at (4.5,2.5)[text=white] {-0.2};
      \node at (4.5,3.5)[text=white] {-0.9};
      \node at (4.5,4.5)[text=white] {0.6};
      \node at (4.5,5.5)[text=white] {0.5};
      
      \node at (5.5,1.5)[text=white] {-0.2};
      \node at (5.5,2.5)[text=white] {-0.2};
      \node at (5.5,3.5)[text=white] {-0.9};
      \node at (5.5,4.5)[text=white] {0.5};
      \node at (5.5,5.5)[text=white] {1.2};
    \end{axis}
    \node at (-0.2,2.2)[text=black] {1};
    \node at (-0.2,1.7)[text=black] {2};
    \node at (-0.2,1.2)[text=black] {3};
    \node at (-0.2,0.7)[text=black] {4};
    \node at (-0.2,0.2)[text=black] {5};
    
    \node at (0.35,-0.2)[text=black] {1};
    \node at (1.05,-0.2)[text=black] {2};
    \node at (1.77,-0.2)[text=black] {3};
    \node at (2.5,-0.2)[text=black] {4};
    \node at (3.25,-0.2)[text=black] {5};
  \end{tikzpicture}

%% file: Tikz/H_inf_red.tex
 \begin{tikzpicture}
    \begin{axis}[view={0}{90},
    grid=both,
    grid style={black!90},
    tick style={black!90},
    width=5.2cm,
    height=4cm,
        y dir=reverse,
        ytick={1,2,...,5},
xtick={1,2,...,5},
yticklabels={,,},
xticklabels={,,}]
      \addplot3[surf, shader=flat corner, scale=0.5] table {Tikz/H_inf_red.txt};
      \node at (1.5,1.5)[text=black] {1.4};
      \node at (1.5,2.5)[text=white] {0.4};
      \node at (1.5,3.5)[text=white] {0.4};
      \node at (1.5,4.5)[text=white] {0.4};
      \node at (1.5,5.5)[text=white] {0.4};
      
      \node at (2.5,1.5)[text=white] {0.4};
      \node at (2.5,2.5)[text=black] {1.4};
      \node at (2.5,3.5)[text=white] {0.4};
      \node at (2.5,4.5)[text=white] {0.6};
      \node at (2.5,5.5)[text=white] {0.4};
      
      \node at (3.5,1.5)[text=white] {0.4};
      \node at (3.5,2.5)[text=white] {0.4};
      \node at (3.5,3.5)[text=black] {1.4};
      \node at (3.5,4.5)[text=white] {0.4};
      \node at (3.5,5.5)[text=white] {0.4};
      
      \node at (4.5,1.5)[text=white] {0.4};
      \node at (4.5,2.5)[text=white] {0.6};
      \node at (4.5,3.5)[text=white] {0.4};
      \node at (4.5,4.5)[text=black] {1.4};
      \node at (4.5,5.5)[text=white] {0.8};
      
      \node at (5.5,1.5)[text=white] {0.4};
      \node at (5.5,2.5)[text=white] {0.4};
      \node at (5.5,3.5)[text=white] {0.4};
      \node at (5.5,4.5)[text=white] {0.8};
      \node at (5.5,5.5)[text=black] {1.4};
    \end{axis}
    \node at (-0.2,2.2)[text=black] {1};
    \node at (-0.2,1.7)[text=black] {2};
    \node at (-0.2,1.2)[text=black] {3};
    \node at (-0.2,0.7)[text=black] {4};
    \node at (-0.2,0.2)[text=black] {5};
    
    \node at (0.35,-0.2)[text=black] {1};
    \node at (1.05,-0.2)[text=black] {2};
    \node at (1.77,-0.2)[text=black] {3};
    \node at (2.5,-0.2)[text=black] {4};
    \node at (3.25,-0.2)[text=black] {5};
  \end{tikzpicture}